\documentclass[twocolumn,english,reprint, longbibliography, superscriptaddress, breaklinks=true, showkeys, showpacs=false, nofootinbib]{revtex4}
\usepackage[T1]{fontenc}
\usepackage[latin9]{inputenc}
\usepackage{color}
\usepackage{babel}
\usepackage{amsmath}
\usepackage{amssymb}
\usepackage{graphicx}
\usepackage{physics}
\usepackage{subfigure}
\usepackage[unicode=true,pdfusetitle,
 bookmarks=true,bookmarksnumbered=false,bookmarksopen=false,
 breaklinks=true,pdfborder={0 0 0},backref=false,colorlinks=true]
 {hyperref} 
\usepackage{breakurl}
\usepackage{qcircuit}

\makeatletter
\@ifundefined{textcolor}{}
{%
 \definecolor{BLACK}{gray}{0}
 \definecolor{WHITE}{gray}{1}
 \definecolor{RED}{rgb}{1,0,0}
 \definecolor{GREEN}{rgb}{0,1,0}
 \definecolor{BLUE}{rgb}{0,0,1}
 \definecolor{CYAN}{cmyk}{1,0,0,0}
 \definecolor{MAGENTA}{cmyk}{0,1,0,0}
 \definecolor{YELLOW}{cmyk}{0,0,1,0}
}

\pdfoutput=1
\hypersetup{colorlinks=true,citecolor=blue,linkcolor=cyan,urlcolor=blue,filecolor= green, breaklinks=true}
\usepackage{url}
\usepackage{breakurl}
\makeatother

\begin{document}

\title{Reality variation under monitoring with weak measurements}

\author{Marcos L. W. Basso}
\email{marcoslwbasso@hotmail.com}
\thanks{corresponding author}
\address{Departamento de F\'isica, Centro de Ci\^encias Naturais e Exatas, Universidade Federal de Santa Maria, Avenida Roraima 1000, Santa Maria, Rio Grande do Sul, 97105-900, Brazil}
\address{New address: Centro de Ci\^encias Naturais e Humanas, Universidade Federal do ABC, Avenida dos Estados 5001, 09210-580 Santo Andr\'e, S\~ao Paulo, Brazil}

\author{Jonas Maziero}
\email{jonas.maziero@ufsm.br}
\address{Departamento de F\'isica, Centro de Ci\^encias Naturais e Exatas, Universidade Federal de Santa Maria, Avenida Roraima 1000, Santa Maria, Rio Grande do Sul, 97105-900, Brazil}

\selectlanguage{english}%

\begin{abstract}
\textbf{Abstract}:
Recently, inspired by Einstein-Podolsky-Rosen's notion of elements of reality, Bilobran and Angelo gave a formal and operational characterization of (ir)reality [EPL 112, 40005 (2015)]. From this approach, the authors were able to define a measure of (ir)realism, or (in)definiteness, of an observable given a preparation of a quantum system. As well, in [Phys. Rev. A 97, 022107 (2018)], Dieguez and Angelo studied the variation of reality of observables by introducing a map, called monitoring, through weak projective non-revealed measurements. The authors showed that an arbitrary-intensity unrevealed measurement of a given observable $X$ generally increases its reality, also increasing the reality of its incompatible observables $X'$. However, from these results, natural questions arise: under the monitoring map of $X$, how much does the reality of $X'$ increase in comparison to that of $X$? Does it always increase? This is the kind of question we address in this article. Surprisingly, we show that it is possible that the variation of the reality of $X'$ is bigger than the variation of the reality of $X$. As well, the monitoring map of $X$ does not affect the already established reality of $X'$, even when they are maximally incompatible. On the other hand, there are circumstances where the variation of reality of both observables is the same, even when they are maximally incompatible. Besides, we give a quantum circuit to implement the monitoring map and use it to experimentally verify the variation of reality of observables using IBM's quantum computers.

\end{abstract}


\maketitle

\section{Introduction}
In 1935, the guidelines for the comprehension of important aspects of the quantum theory were mainly given by Bohr's ideas about complementarity \cite{Bohr} as well as by Heisenberg's uncertainty principle \cite{Robertson}. These concepts were basically concerned with observation and measurement in the quantum realm. However, in May of the same year, Einstein, Podolsky, and Rosen (EPR) published a seminal article criticizing the conceptual understanding of quantum theory \cite{Einstein}, more specifically, about its completeness. Since then, this paper became a cornerstone in the discussions about the foundations of quantum mechanics, having an important role in the development of quantum information theory as well. The work of EPR starts arguing that every physical theory must be complete. For this end, EPR defined the notion of element of reality: 
``If, without in any way disturbing a system, we can predict with certainty the value of a physical quantity, then there exists an element of physical reality corresponding to this physical quantity.'' Therefore, ``every
element of the physical reality must have a counter part in the physical theory'', which they called the
condition of completeness. Afterwards, they considered a case study where two quantum systems interact with each other such that they end up quantum correlated, even when the systems are widely separated in space. This quantum correlation is by now well understood and it is known as entanglement. Together with the notion of locality, the authors disproved the completeness of quantum theory by arguing that these ingredients imply that non-commuting observables can be simultaneously determined. 

However, as noticed by Bell \cite{Bell}, and confirmed by loopholes-free experiments \cite{Alain, Aspect, Hensen, Shalm}, any theory aiming at completing quantum mechanics cannot be fulfilled with hidden local causal variables. For instance, Bohmian mechanics is a realistic hidden-variable theory where local causality must be discarded \cite{Bohm}. On the other hand, one can always drop the pre-defined notion of reality (or definiteness) of an observable, so attaining the notion of local causality. Besides, more recently, an alternative attitude toward Bell's theorem has been developed, inspired by the framework of causal inference. In this approach, Bell's inequality violation does not lead to the quandary between realism and local causality. Instead, it attests to the impossibility of providing a non-fine-tuned explanation of the experiment within the framework of classical causal models \cite{Spekkens, Wolfe}. Independent of these notions, it is well known nowadays that quantum mechanics does not allow for instantaneous communication at distance, what Einstein called ``spooky action at distance''. It is actually not difficult to realize that standard quantum mechanics is a local theory in this sense \cite{Terno}. Nevertheless, the violation of Bell's inequalities leaves no doubt that the classical deterministic notion of the reality of an observable or local causality  deserves a meticulous examination. In this work, we follow the line of research of reviewing the notion of realism, as the authors in Refs. \cite{Renato, Dieguez}.

Recently, inspired by EPR's elements of reality \cite{Einstein}, Bilobran and Angelo \cite{Renato} reported a formal operational notion of (ir)reality. From this approach, they were able to define a measure of (ir)realism, or (in)definiteness, of an observable given a preparation of a quantum system. This definition, together with the measure of realism, have been proven fruitful in several contexts \cite{Dieguez, Gomes, Angelo, Fucci, Orthey, Moreira, Marcos, Ale, Serra}. For instance, in \cite{Ale}, the authors built an axiomatization for the notion of quantum realism inspired by the fact that the encoding of information about a given observable in a physical degree freedom is a necessary condition for such an observable to become an element of the physical reality. While, in \cite{Serra}, the authors considered an operational criterion of physical reality for the wave-particle aspect of a quantum system and provided a setup that ensures a formal link between the output visibility and elements of reality within the interferometer. 

Besides, in Ref. \cite{Dieguez} the authors employed this measure in order to establish relations among the concepts of measurement, information, and physical reality. As well, after introducing a map called monitoring through weak projective non-revealed measurements, the authors were able to show that an arbitrary-intensity unrevealed measurement of a given observable $X$ generally leads to an increase of its reality and also of the reality of  observables $X'$ incompatible to $X$, which is an important result regarding the emergence of the classical world from the quantum realm \cite{Zurek, Horo, Korbicz}. However, from these results, several questions arise: how much does the reality of the observable $X'$ increase in comparison with the reality of the observable $X$ through the monitoring map of $X$? Does it always increase? If no, under what  conditions does the reality of $X'$ increase? Is it possible that the reality of the observable $X'$ increases more than the reality of the observable $X$ under monitoring of $X$? This is the kind of questions that we address in this article. Besides, we give a quantum circuit to implement the monitoring map on quantum computers and we experimentally verify the variation of reality of observables using IBM's quantum computers. 

The remainder of this article is organized as follows. In Sec. \ref{sec:elre}, we review the framework developed by Bilobran and Angelo and discuss the variation of reality under weak non-revealed measurements. In Sec. \ref{sec:var}, we present our main results by answering the questions raised above. In Sec. \ref{sec:exp}, we present a quantum circuit to implement the monitoring map through weak non-revealed measurements and experimentally verify the variation of reality for some states and observables, that are used to answer the questions raised in this work. Finally, in Sec. \ref{sec:con}, we give our concluding remarks.

%-----------------------
\section{Elements of reality}
\label{sec:elre}

In this section, we review the framework put forward by Bilobran and Angelo in Ref. \cite{Renato}, and we discuss the variation of reality of observables under the monitoring map through weak unrevealed measurements, as introduced in Ref. \cite{Dieguez}. First, let us consider a preparation $\rho$ of a quantum system $A$. Second, it is performed, between the preparation and the tomography procedures, a non-selective projective measurement of an observable $X$, where $X = \sum_j x_j \Pi^{X}_j$  is a discrete spectrum observable, with  $\Pi^{X}_j = \ketbra{x_j}{x_j}$ being orthonormal projectors acting  on the Hilbert space $\mathcal{H}_A$ of the quantum system $A$. Since no information about the measurement outcomes is revealed, the post measurement state is given by \cite{Groenewold, Busch}:
\begin{align}
    \Phi_{X}(\rho) =  \sum_j  \Pi^{X}_j \rho \Pi^{X}_j.
\end{align}
The next step is to compare the preparation $\rho$ with the state of the system after the non-revealed measurements, i.e., $\Phi_X(\rho)$. When $\rho = \Pi^{X}_j$, for some $j$, the observer can conclude that an element of reality for $X$ was already implied in the preparation, which agrees with EPR's notion of reality of the observable $X$. However, it also predicts an element of reality for $\rho = \sum_j p^X_j \Pi^{X}_j = \Phi_X (\rho)$, where $p^X_j = \Tr (\rho \Pi^{X}_j)$. Therefore, the operational definition given by Bilobran and Angelo generalizes the notion of reality of an observable first introduced by EPR. So, the authors in Ref. \cite{Renato} raised the procedure of non-revealed measurements as the main ingredient for establishing the reality of the observable $X$ given the preparation $\rho$. 

With this in mind, they also defined the following measure of local irreality (or indefiniteness) of $X$ given $\rho$: 
\begin{equation}
\mathfrak{I}_X(\rho):= S(\Phi_{X}(\rho)) - S(\rho), \label{eq:locirre}
\end{equation}
where $S(\rho) = -\Tr \rho \log \rho$ is the von Neumann entropy and $\log = \log_2$. Eq. (\ref{eq:locirre}) already appear in the literature in different forms and contexts with different interpretations. For instance, in the context of average information gain by quantum measurements \cite{Groenewold, Busch} and it's directly related to the quantum coherence based on the relative entropy \cite{Baum}. Besides, it is worth mentioning that, in Ref. \cite{Renato}, the authors first introduced the notion of irrealism of an observable $X$ for bipartite quantum systems. Here, we shall deal only with local irreality of the observable $X$. From this,  it is straightforward to define the local reality (or definiteness) of the observable $X$, given the preparation $\rho$, as
\begin{align}
    \mathfrak{R}_X(\rho) & := \log d - \mathfrak{I}_X(\rho).
\end{align}
It is noteworthy that in Ref. \cite{Marcos} we established connections between the measures introduced by Bilobran and Angelo with the measures that quantify the complementarity properties of a quantum system, including its entanglement with other quantum systems. For instance, the local reality of the observable $X$ is related to the predictability measure of the observable $X$ before a projective measurement, i.e., its ``pre-existing'' reality as well as the possible generation of entanglement with an informer, i.e., a degree of freedom that records the information about the state of the system, while the irreality of $X$ is directly related to the quantum coherence of $\rho$ in the eigenbasis of $X$, as already noticed in Ref. \cite{Renato}.

To introduce the notion of the monitoring through weak unrevealed measurements, let us consider another quantum system $B$, called ancilla, which will couple to our quantum system $A$ in order to encode the information about $A$. Basically, the authors used the Stinespring's dilation theorem \cite{Paris} to model such monitoring map. By considering a initial separable state $\rho_{AB} = \rho \otimes \ketbra{b}{b}$, where $\ketbra{b}{b}$ is the initial state of ancilla, and under a suitable global unitary evolution operator $U$, the authors showed that
\begin{align}
    \Phi^{\epsilon}_X(\rho)  =  \Tr_B (U \rho \otimes \ketbra{b}{b} U^{\dagger}) =  (1 - \epsilon) \rho + \epsilon \Phi_X (\rho),
\end{align}
with $\epsilon \in [0,1]$. One can readily see that the map $\Phi^{\epsilon}_X$ interpolates continuously between no measurement at all ($\epsilon = 0$) and a strong projective non-revealed measurement ($\epsilon = 1$).

Now, given the preparation $\rho$ of our system $A$, under the monitoring $\Phi_X^{\epsilon}$ of arbitrary intensity $\epsilon$, the initial reality of the observable $X$, given by $\mathfrak{R}_X(\rho)$, will change to $\mathfrak{R}_X(\Phi_X^{\epsilon}(\rho))$. Therefore, the variation of the reality of the observable $X$ under the monitoring map is given by
\begin{align}
    \Delta \mathfrak{R}_X & = \mathfrak{R}_X\Big(\Phi_X^{\epsilon}(\rho)\Big) - \mathfrak{R}_X(\rho) \nonumber \\ & = S(\Phi_X^{\epsilon}(\rho)) - S(\rho), \label{eq:varx}
\end{align}
which is a non-negative quantity. Besides, the authors in Ref. \cite{Dieguez} showed that $\Delta \mathfrak{R}_X \ge \epsilon \mathfrak{I}_X(\rho)$, an inequality that was verified experimentally in Ref. \cite{Pater}.

Next, given another observable $X'$, which can be incompatible with $X$, the authors in Ref. \cite{Dieguez} asked the following question: how much does the reality of $X'$ vary when a monitoring $\Phi_X^{\epsilon}$ is performed on $\rho$? Given that the initial reality of $X'$ is $\mathfrak{R}_{X'}(\rho)$ and, under the monitoring map $\Phi_X^{\epsilon}$, the reality of $X'$ changes to $\mathfrak{R}_{X'}(\Phi_X^{\epsilon}(\rho))$. So
\begin{align}
    \Delta \mathfrak{R}_{X'} & = \mathfrak{R}_{X'}(\Phi_X^{\epsilon}(\rho))- \mathfrak{R}_{X'}(\rho) \label{eq:varx'}\\
    & = S(\Phi_{X'}(\rho)) + S(\Phi^{\epsilon}_X(\rho)) -  S(\rho) - S(\Phi_{X'} \Phi^{\epsilon}_X(\rho)). \nonumber
\end{align}
By using the strong sub-additivity of the von Neumann entropy, the authors managed to show that $\Delta \mathfrak{R}_{X'} \ge 0$, which, at first, seems an unexpected result, i.e., the variation of reality  of $X'$ never decreases under the monitoring map of $X$. From now on, one of our main goals is to compare the variations $\Delta\mathfrak{R}_{X'}$ and $\Delta\mathfrak{R}_{X}$.

%--------------------------------
\section{Variation of realities}
\label{sec:var}
In this section, we will compare $\Delta\mathfrak{R}_{X'}$ with $\Delta\mathfrak{R}_{X}$, and we shall address the following questions: through the monitoring map of $X$, how much does the reality of the observable $X'$ increase in comparison with the reality of the observable $X$? Does it always increase? If no, under what conditions does the reality of $X'$ increase? Is it possible that the reality of the observable $X'$ increases more or equally to the reality of the observable $X$? Besides, it's worth mentioning that the results obtained in this section remains valid for the general scenario where we have a bipartite quantum system and $X$ is an observable of one of the parts of the quantum system.

First, let us notice that, from Eqs. (\ref{eq:varx}) and (\ref{eq:varx'}), we have
\begin{align}
    \Delta\mathfrak{R}_{X'} = \Delta\mathfrak{R}_{X} +  S(\Phi_{X'}(\rho)) - S(\Phi_{X'} \Phi^{\epsilon}_X(\rho)). \label{eq:varreal}
\end{align}
From the equation above, we can see that: \textit{(i)} if $X$ and $X'$ are compatible observables, i.e., if $[X,X'] = 0$, then it is easy to show that $\Phi_{X'}(\rho) = \Phi_{X'} \Phi^{\epsilon}_X(\rho)$, which implies that $\Delta\mathfrak{R}_{X'} = \Delta\mathfrak{R}_{X}$. Therefore, under the monitoring map $\Phi^{\epsilon}_X$, the variation of the realities of $X$ and $X'$ are the same when they are compatible; \textit{(ii)} if the initial state $\rho$ is prepared in an eigenstate of $X$, or more generally, in a mixture of eigenstates of $X$, i.e, $\rho = \sum_j p^X_j \Pi^X_j$, then $\Phi_X^{\epsilon}(\rho) = \rho$ and $\Phi_{X'}(\rho) = \Phi_{X'}(\Phi_X^{\epsilon}(\rho))$, which implies that $\Delta\mathfrak{R}_{X'} = \Delta\mathfrak{R}_{X} = 0$. Here, we \textit{do not} assume that $X$ and $X'$ are compatible. This means that, given the established reality of $X$, the variation of reality of any other observable $X'$ is null under the monitoring map $\Phi_X^{\epsilon}$. Therefore, one can see that the prepared state and the choice of the monitoring map control the variation of the reality of any other observable, such that the necessary condition for the reality of $X'$  to change under the monitoring of $X$ is the state preparation to be different from $\rho = \sum_j p^X_j \Pi^X_j$. However, as we will see in the cases below, this is not a sufficient condition.

Now, let us consider more interesting cases: \textit{(iii)}  If the initial state is prepared in an eigenbasis of $X'$ or, more generally, in a mixture of eigenstates of $X'$, i.e., $\rho = \sum_j p^{X'}_j \Pi^{X'}_j$ where $p^{X'}_j = \Tr (\rho \Pi^{X'}_j)$ and $\Pi^{X'}_j = \ketbra{x'_j}$, then $\Delta \mathfrak{R}_{X'} = 0$. This a direct consequence of $\Delta \mathfrak{R}_{X'} \ge 0$ \cite{Dieguez} and, since the reality of $X'$ is already established in the preparation (i.e., $\mathfrak{R}_{X'}(\rho) =  \mathfrak{R}^{\max}_{X'}$), we have $\Delta \mathfrak{R}_{X'} = \mathfrak{R}_{X'}(\Phi^{\epsilon}_{X}(\rho)) -\mathfrak{R}^{\max}_{X'} \le 0$, leaving the only option $\Delta \mathfrak{R}_{X'} = 0$. Even though it is mathematically trivial, this is an interesting result, i.e, the monitoring of the observable $X$ does not affect the already established reality of the observable $X'$, even if they are maximally incompatible. 

To illustrate the case of maximally incompatible observables more clearly, let $X$ and $X'$ be maximally incompatible observables, i.e., $[X,X'] \neq 0$ and their eigenbasis are mutually unbiased (MU), i.e., $\abs{\braket{x_j}{x'_k}}^2 = 1/d, \ \ \forall j,k$, where $\{\ket{x_j}\}$ and $\{\ket{x'_k}\}$ are the eigenbasis of $X$ and $X'$ respectively, and $d := \dim \mathcal{H}_A$ is the dimension of the Hilbert space of the system A. Thus
\begin{align}
& \Phi^{\epsilon}_X(\rho) =  (1 - \epsilon) \rho  + \epsilon \Phi_X(\rho), \label{eq:phie} \\
& \Phi_{X'}(\Phi^{\epsilon}_X(\rho)) = (1 - \epsilon) \Phi_{X'}(\rho) + \epsilon I/d, \label{eq:phix},
\end{align}
where $I$ is the identity matrix. If the initial state is given by $\rho = \sum_j p^{X'}_j \Pi^{X'}_j$, then this it is enough to realize that Eqs. (\ref{eq:phie}) and (\ref{eq:phix}) are the same. Therefore, we have $S(\Phi^{\epsilon}_X(\rho)) = S(\Phi_{X'} \Phi^{\epsilon}_X(\rho))$ and $S(\Phi_{X'}(\rho)) = S(\rho)$, which proves that $\Delta \mathfrak{R}_{X'} = 0$. However, this result does not imply that the variation of the reality of $X$ is null, since $S(\Phi^{\epsilon}_X(\rho))$ can be different from $S(\rho)$. On the other hand, if \textit{(iv)} $\rho$ is not an eigenstate of $X'$  neither a mixture of its eigenstates, then, by the concavity of the von Neumann entropy, we have
\begin{align}
S(\Phi_{X'} \Phi^{\epsilon}_X(\rho)) - S(\Phi_{X'}(\rho))\nonumber & \ge \epsilon(\log d - S(\Phi_{X'}(\rho))) \\& \ge 0,    
\end{align}
which implies that $\Delta \mathfrak{R}_X \ge \Delta \frak{R}_{X'}$ when $X$ and $X'$ are maximally incompatible observables. 

Another interesting case is the following one: \textit{(v)} the initial state of system is prepared in an eigenstate of the observable $X''$, or more generally, in the state $\rho = \sum_j p^{X''}_j \Pi^{X''}_j$ where $p^{X''}_j = \Tr (\rho \Pi^{X''}_j)$ and $\Pi^{X''}_j = \ketbra{x''_j}$, and $X, X', X''$ is a set of maximally incompatible observables. The already established reality of $X''$ is not affected under the monitoring of $X$ (or of $X'$), as already discussed in the  case $\textit{(iii)}$. Besides that, it is possible to show that $\Delta \mathfrak{R}_{X'} = \Delta \mathfrak{R}_{X} \neq 0$, i.e., the variation of reality of the maximally incompatible observables $X$ and $X'$ are the same. To see this, let us consider the eigenstates $\{|x_j\rangle, |x'_j\rangle, |x''_j\rangle\}$ of $X, X'$ and $X''$, respectively, such that
\begin{align}
|\langle x_j | x'_k \rangle|^2 = |\langle x_j | x''_k \rangle|^2 = |\langle x'_j | x''_k \rangle|^2 = 1/d, \ \ \ \forall j,k.
\end{align}
Now, if the prepared state $\rho$ is $\ketbra{x''_j}$ for some $j$ or $\rho = \sum_j p^{X''}_j \Pi^{X''}_j$, then
\begin{align}
& \Phi^{\epsilon}_X (\rho) = (1 - \epsilon) \rho + \epsilon I/d, \\
& \Phi_{X'}\Big(\Phi^{\epsilon}_X (\rho)\Big) = (1 - \epsilon) I/d + \epsilon I/d = I/d, \\ 
& \Phi_{X'} (\rho) = I/d.
\end{align}
Therefore $S_{vn}\Big(\Phi_{X'}\Phi^{\epsilon}_X (\rho)\Big) = S_{vn}\Big(\Phi_{X'}(\rho)\Big) $, which implies that
\begin{align}
\Delta \mathfrak{R}_{X'} = \Delta \mathfrak{R}_{X} \neq 0. 
\end{align}

Lastly, from Eq. (\ref{eq:varreal}), one can see that $\Delta \mathfrak{R}_{X'} \gtrless  \Delta \mathfrak{R}_{X}$ when $S(\Phi_{X'}(\rho)) \gtrless S(\Phi_{X'} \Phi^{\epsilon}_X(\rho))$. In order to show that both situations are possible, we will give examples using qubits. First, let us assume that we prepare a qubit in the state $|\psi \rangle = |+ \rangle = \frac{1}{\sqrt{2}}(\ket{0} + \ket{1})$. So $S(\rho=|\psi\rangle\langle\psi|) = 0$. The monitoring map is applied using the observable $X = \sigma_z$, which is one of the Pauli matrices and whose eigenstates are $|0\rangle, |1\rangle$. The observable $X'$ is $\hat{n} \cdot \vec{\sigma}$, where $\vec{\sigma} = (\sigma_x, \sigma_y, \sigma_z)$ are the Pauli matrices and $\hat{n}$ is a unit vector of $\mathbb{R}^{3}$. The eigenvectors of $X' = \hat{n} \cdot \vec{\sigma}$ are $|n_0\rangle, |n_1\rangle$, where
\begin{align}
& |n_{0}\rangle = \cos(\theta/2)|0\rangle + e^{i\phi} \sin(\theta/2)|1\rangle, \\
&|n_{1}\rangle = -\sin(\theta/2)|0\rangle + e^{i\phi}\cos(\theta/2)|1\rangle.
\end{align}
\begin{figure}[t]
    \centering
    \includegraphics[scale=0.5]{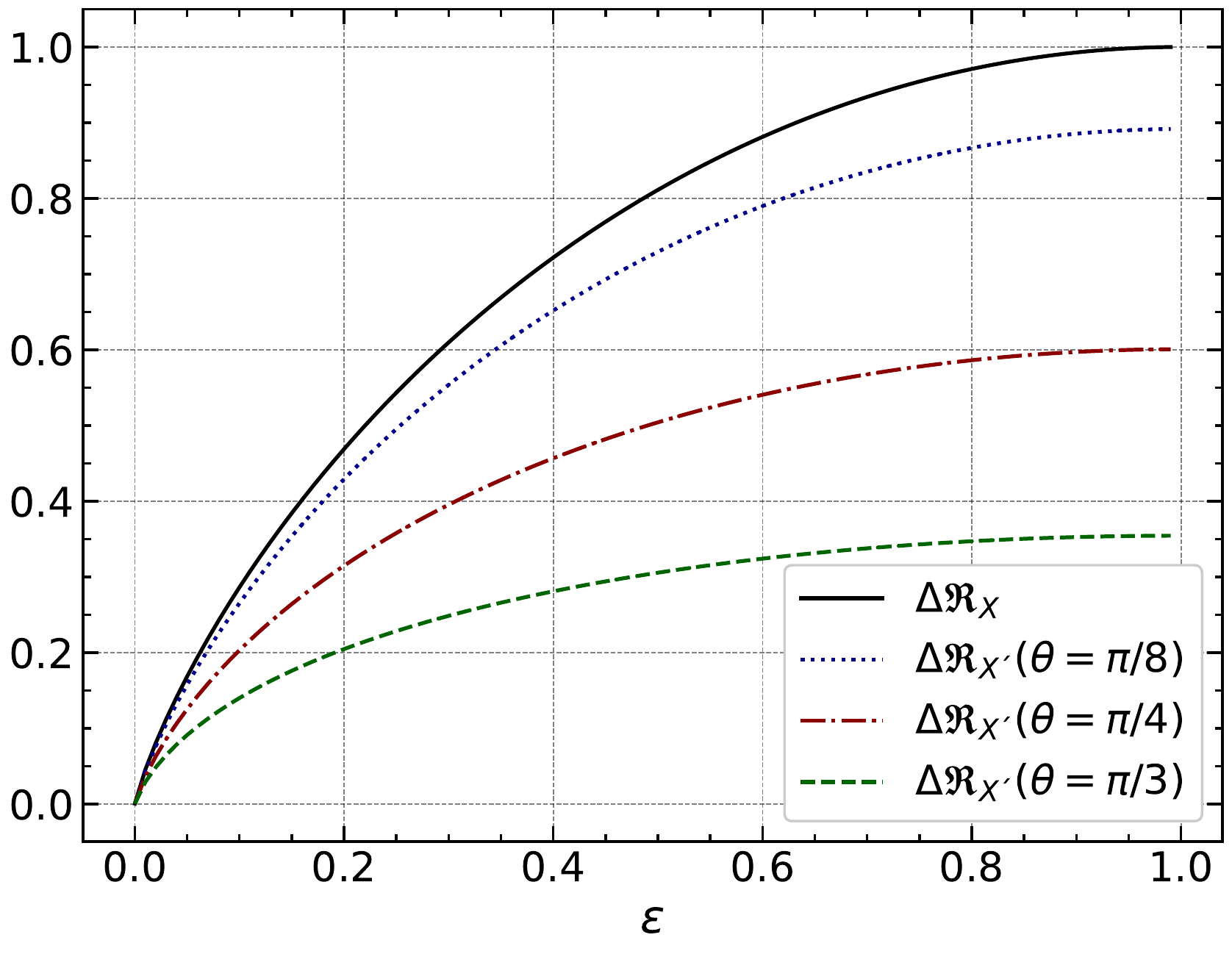}
    \caption{Comparison between  $\Delta \mathfrak{R}_{X'}$ and $\Delta \mathfrak{R}_{X}$ for $\rho=|+\rangle\langle+|$, $X =  \sigma_z$ and  $X' = \hat{n} \cdot \vec{\sigma}$, for different values of $\theta$.}
    \label{fig:varrea}
\end{figure}
Hereafter we set $\phi=0$. In this case, we have the post-measurement states
\begin{align}
& \Phi^{\epsilon}_{X}(\rho) = (1 - \epsilon) \rho + \epsilon \frac{I_{2 \times 2}}{2} ,\\
& \Phi_{X'}(\rho) = \frac{1}{2}\sum_{j = 0}^1( 1 + (-1)^j \sin \theta) \Pi^{X'}_{j}, \\
& \Phi_{X'}\Big(\Phi^{\epsilon}_{X}(\rho)\Big) = \frac{1 - \epsilon}{2}\sum_{j = 0}^1( 1 + (-1)^j \sin \theta) \Pi^{X'}_{j}  + \epsilon \frac{I_{2 \times 2}}{2} ,
\end{align}
whose eigenvalues are given, respectively, by
\begin{align}
& \lambda^X_{\pm} = \frac{1}{2}( 1 \pm (1 -\epsilon) ), \\
& \lambda^{X'}_{\pm} = \frac{1}{2}( 1 \pm \sin \theta), \\
& \lambda^{XX'}_{\pm} = \frac{1}{2}( 1 \pm (1 - \epsilon) \sin \theta).
\end{align} 
From these eigenvalues, it is possible to calculate $\Delta \mathfrak{R}_{X'}$ and $\Delta \mathfrak{R}_{X}$ for comparison. As one can see in Fig. \ref{fig:varrea}, in this case, we have $\Delta \mathfrak{R}_{X} \ge \Delta \mathfrak{R}_{X'}$ for $\theta \in (0, \pi)$.

Now, let us consider that the monitoring map is given by the observable $X = \hat{n} \cdot \vec{\sigma}$, while $X' = \sigma_z$. If the system's prepared state is $| \psi \rangle = | + \rangle$, then
\begin{align}
& \Phi^{\epsilon}_{X}(\rho) = (1 - \epsilon) \rho + \frac{\epsilon}{2}\sum_{j = 0}^1( 1 + (-1)^j \sin \theta) \Pi^{X'}_{j},\\
& \Phi_{X'} (\rho) = \frac{I_{2 \times 2}}{2},\\
& \Phi_{X'}\Big(\Phi^{\epsilon}_{X}(\rho \Big) =  \frac{I_{2 \times 2}}{2} + \frac{\epsilon}{2} \cos \theta \sin \theta \sigma_z.
\end{align}
The last two density operators are diagonal, therefore their eigenvalues are straightforward to obtain, while the eigenvalues of $\Phi^{\epsilon}_{X}(\rho)$ are given by
\begin{align}
    \lambda_{\pm} = \frac{1}{2}\Big(1 \pm \sqrt{\epsilon^2\sin^2\theta\cos^2\theta + (1 - \epsilon\cos^2\theta)^2}\Big).
\end{align}

\begin{figure}[t]
    \centering
    \includegraphics[scale=0.5]{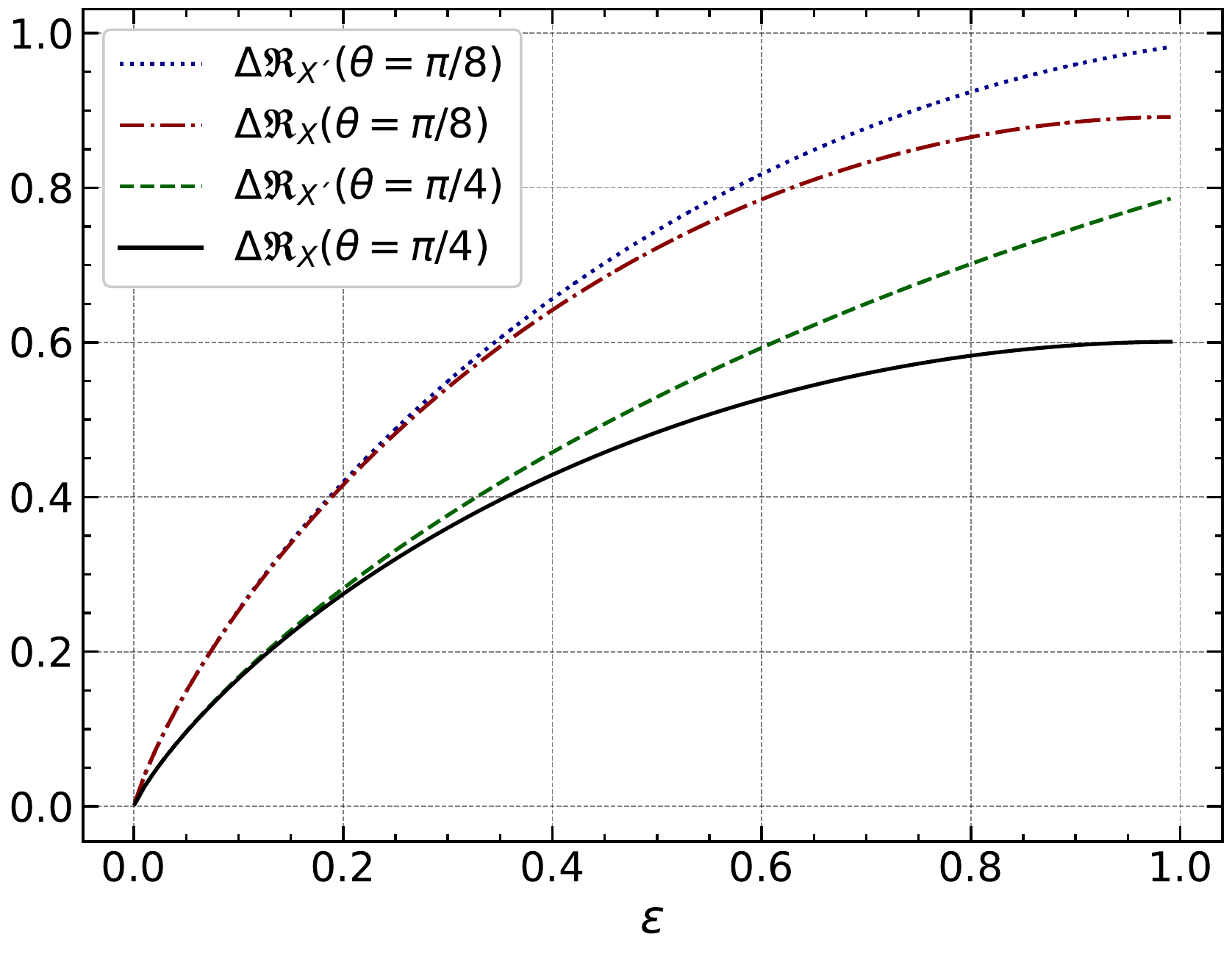}
    \caption{Comparison between  $\Delta \mathfrak{R}_{X'}$ and $\Delta \mathfrak{R}_{X}$ for $X = \hat{n} \cdot \vec{\sigma}$ and $X' =  \sigma_z$, for different values of $\theta$.}
    \label{fig:varrea1}
\end{figure}

Thus, in this case, the variation of reality of $X$ will also depend on $\theta$, as one can see in Fig. \ref{fig:varrea1}. Besides, here we have $\Delta \mathfrak{R}_{X'} \ge \Delta \mathfrak{R}_{X}$ for any $\theta \in (0, \pi)$. Since $\Phi_{X'} (\rho) = \frac{I_{2 \times 2}}{2}$, then $S(\Phi_{X'} (\rho)) = 1 = S^{\max}$, which implies that
\begin{align}
     \Delta\mathfrak{R}_{X'} & = \Delta\mathfrak{R}_{X} +  S^{\max} - S(\Phi_{X'} \Phi^{\epsilon}_X(\rho))\\
     & \ge \Delta\mathfrak{R}_{X}. \nonumber
\end{align}
Therefore, with this we show the interesting, and unexpected, case where it is possible that the variation of reality of an observable $X'$ is bigger than the variation of reality of the observable $X$ under the monitoring of $X$.

%------------------------
\section{Variation of reality in the IBM Quantum Experience }
\label{sec:exp}

In this section, we shall use IBM's quantum computers \cite{ibmq} to verify experimentally the  variation of reality of the observables $X$ and $X'$, as discussed in the previous section. We will start by giving a unitary quantum circuit to perform the monitoring map, i.e., the weak non-revealed measurements on a qubit. The generalization of this quantum circuit for an arbitrary number of qubits is fairly straightforward. This quantum circuit is interesting in its own, since it can be used in different contexts, as for example for experimental tests related to the weak quantum discord defined in Ref. \cite{prd}.

\begin{figure}[t]
%\centering
$\Qcircuit @C=1em @R=.7em { 
\lstick{\rho} & \gate{U^{\dagger}(\theta,\phi,0)} & \ctrl{1} & \gate{U(\theta,\phi,0)} & \qw & \rstick{\Phi^{\epsilon}_{n_{0},n_{1}}(\rho)} \\
\lstick{\ket{0}} & \gate{U(\theta,0,0)}  & \targ & \qw & \qw
}$
\caption{Quantum circuit for implementing an one-qubit weak non-revealing  measurement on a quantum computer. We use $V=U(\theta,\phi,\lambda)$ with $\lambda=0$ and $U^{\dagger}(\theta,\phi,\lambda)=U(\theta,\pi-\lambda,-\pi-\phi)$  with $U(\theta,\phi,\lambda)=\begin{bmatrix} \cos(\theta/2) & -e^{i\lambda}\sin(\theta/2) \\ e^{i\phi}\sin(\theta/2) & e^{i(\phi+\lambda)}\cos(\theta/2) \end{bmatrix}$.}
\label{fig:1qbqc}
\end{figure}
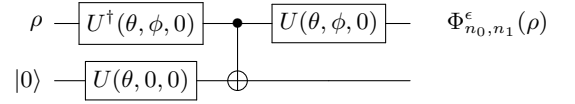

We want to perform a weak non-selective von Neumann measurement of a general one-qubit observable $\hat{n}\cdot\vec{\sigma}$ using joint unitary operations on this qubit and on an auxiliary system, i.e., we want to apply the Stinespring dilation theorem. 
Let us write the eigenbasis of this observable as 
\begin{align}
& |n_{0}\rangle = \cos(\theta/2)|0\rangle + e^{i\phi}\sin(\theta/2)|1\rangle = V|0\rangle, \\
& |n_{1}\rangle = -\sin(\theta/2)|0\rangle + e^{i\phi}\cos(\theta/2)|1\rangle = V|1\rangle,
\end{align}
with $V = U(\theta, \phi, 0) =  \begin{bmatrix} \cos(\theta/2) & -\sin(\theta/2) \\ e^{i\phi}\sin(\theta/2) & e^{i\phi}\cos(\theta/2) \end{bmatrix}$.
First, we give a quantum circuit to implement the monitoring map on the computational basis, i.e., with respect to the observable $X = \sigma_z$. Given the bipartite quantum state
\begin{equation}
|\Psi \rangle_{AB} = |\psi\rangle_A \otimes |\phi\rangle_B =  |\psi\rangle_A \otimes \Big(\cos\frac{\theta}{2} |0 \rangle_B + \sin \frac{\theta}{2} |1 \rangle_B\Big), 
\end{equation}
where $|\phi\rangle_B$ is the state of the ancilla, that can be obtained from $|\phi\rangle_B = U(\theta, 0, 0) |0 \rangle_B$. Following hints from Refs. \cite{marcosPR, Pater}, we apply the controlled-phase operation
\begin{equation}
C_Z (A \to B) := |0 \rangle \langle 0| \otimes I_{2 \times 2} + |1 \rangle \langle 1| \otimes \sigma_z
\end{equation}
to $|\Psi \rangle_{AB}$. Thus, taking the partial trace, we have
\begin{align}
    \Phi^{\epsilon}_{0,1}( |\psi\rangle_A \langle \psi |) & = 
Tr_B\Big(C_Z(A \to B)|\Psi \rangle_{AB} \langle \Psi | C_Z^{\dagger}(A \to B) \Big) \nonumber \\ & =\cos \theta| \psi \rangle \langle \psi| + (1 - \cos \theta) \Pi_{0,1}(| \psi \rangle \langle \psi|),
\end{align}
where $1 - \epsilon = \cos \theta$ with $\theta \in [0, \pi/2]$. So $\epsilon \in [0,1]$. Here, the observable $\Pi_{0,1} = \Phi_{0,1}$, where $Tr_{B}$ is the partial trace operation \cite{ptrace}. Then, it is easy to realize that $\Pi_{n_{0},n_{1}}(|\psi\rangle) = \sum_{j=0}^{1}\Pi_{n_{j}}|\psi\rangle\langle\psi|\Pi_{n_{j}} = V\Pi_{0,1}(\ketbra{\phi})V^{\dagger}$ with $\Pi_{n_{j}} = |n_{j}\rangle\langle n_{j}|$ and $|\phi\rangle = V^{\dagger}|\psi\rangle$. So, by the linearity of quantum dynamics, we see that weak non-selective measurements of a qubit observable $\hat{n}\cdot\vec{\sigma}$, of a system prepared in the state $\rho$, can be implemented using the quantum circuit shown in Fig. \ref{fig:1qbqc}. We also notice that if instead of applying $C_Z (A \to B) $ one applies the Control-NOT operation ($C_{X}(A\rightarrow B) = |0\rangle\langle 0|\otimes I_{2\times 2}+|1\rangle\langle 1|\otimes\sigma_{x}
$), then $\epsilon = 1 - \frac{1}{2} \sin \theta \in [0,1/2]$ for $\theta \in [0, \pi/2]$.

\begin{table}
\caption{\label{tb:belem} Parameters for the ibmq\textunderscore belem chip.}
\begin{tabular}{l c c c}
\hline 
Calibration parameters & Q0 & Q1 & Q2 \tabularnewline
\hline 
\hline 
Frequency (GHz) & 5.09 & 55.97 & 5.36 \tabularnewline
T1 ($\mu$s) & 55.97 & 104.56 & 82.60 \tabularnewline
T2 ($\mu$s) & 94.58 & 117.04 & 62.68 \tabularnewline
Readout error ($10^{-2}$) & 2.08 & 1.92 & 2.13 \tabularnewline
%CNOT error ($10^{-2}$) & $0\_1:1.938$ & $1\_0:1.938$ & 0
%CNOT error ($10^{-2}$) & $0\_2:1.938$ & $1\_0:1.938$ & 0
%\tabularnewline
\hline
\end{tabular}
\end{table}

\begin{figure}[t]
    \centering
    \subfigure[Comparison between  $\Delta \mathfrak{R}_{X'}$ and $\Delta \mathfrak{R}_{X}$ for $X = \sigma_z$ and $X' =  \sigma_x$ with the initial state $\ket{\psi} = \ket{+}$ where $\epsilon = 1 - \cos \theta$.]{{\includegraphics[scale = 0.53]{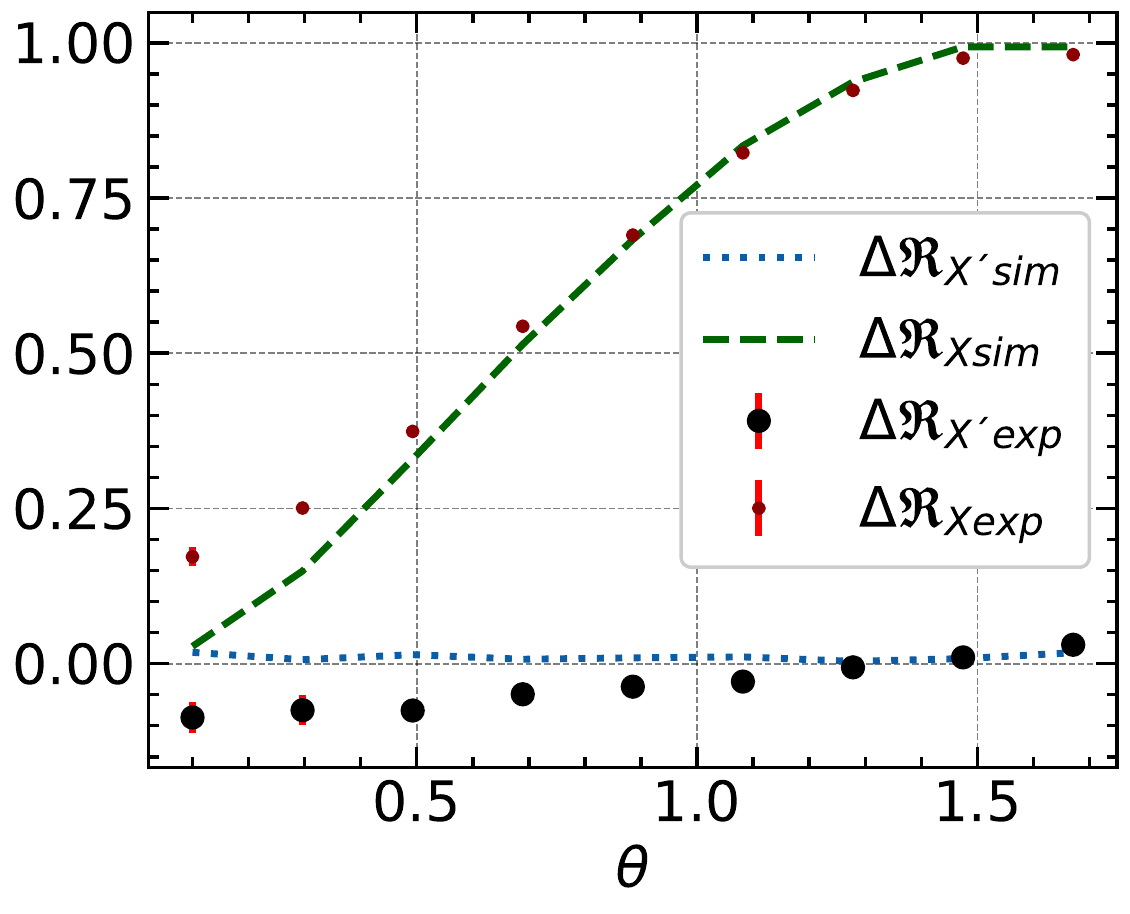}{\label{fig:a}} }}
    \subfigure[Comparison between  $\Delta \mathfrak{R}_{X'}$ and $\Delta \mathfrak{R}_{X}$ for $X = \sigma_z$ and $X' =  \sigma_x$ with the initial state $\ket{\psi} = \frac{1}{\sqrt{2}}(\ket{0} + i \ket{1})$, where $\epsilon = 1 - \cos \theta$.]{{\includegraphics[scale = 0.53]{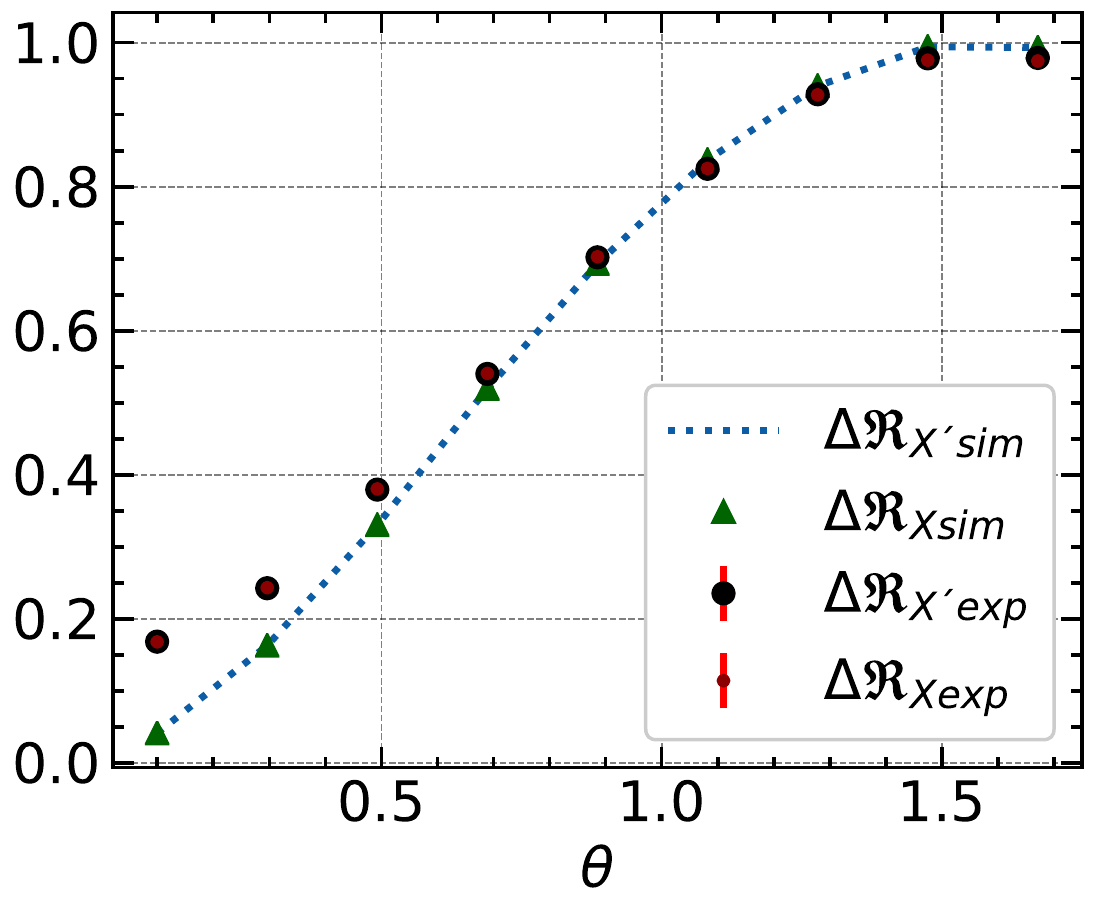}{\label{fig:b}} }}
    \subfigure[An example of $\Delta \mathfrak{R}_{X'} \ge \Delta \mathfrak{R}_{X}$ where the initial state was prepared in $\ket{\psi} = \ket{+}$ and the monitoring map is given by $X - \hat{n} \cdot \vec{\sigma}$ while $X' = \sigma_z$.  
    ]{{\includegraphics[scale = 0.53]{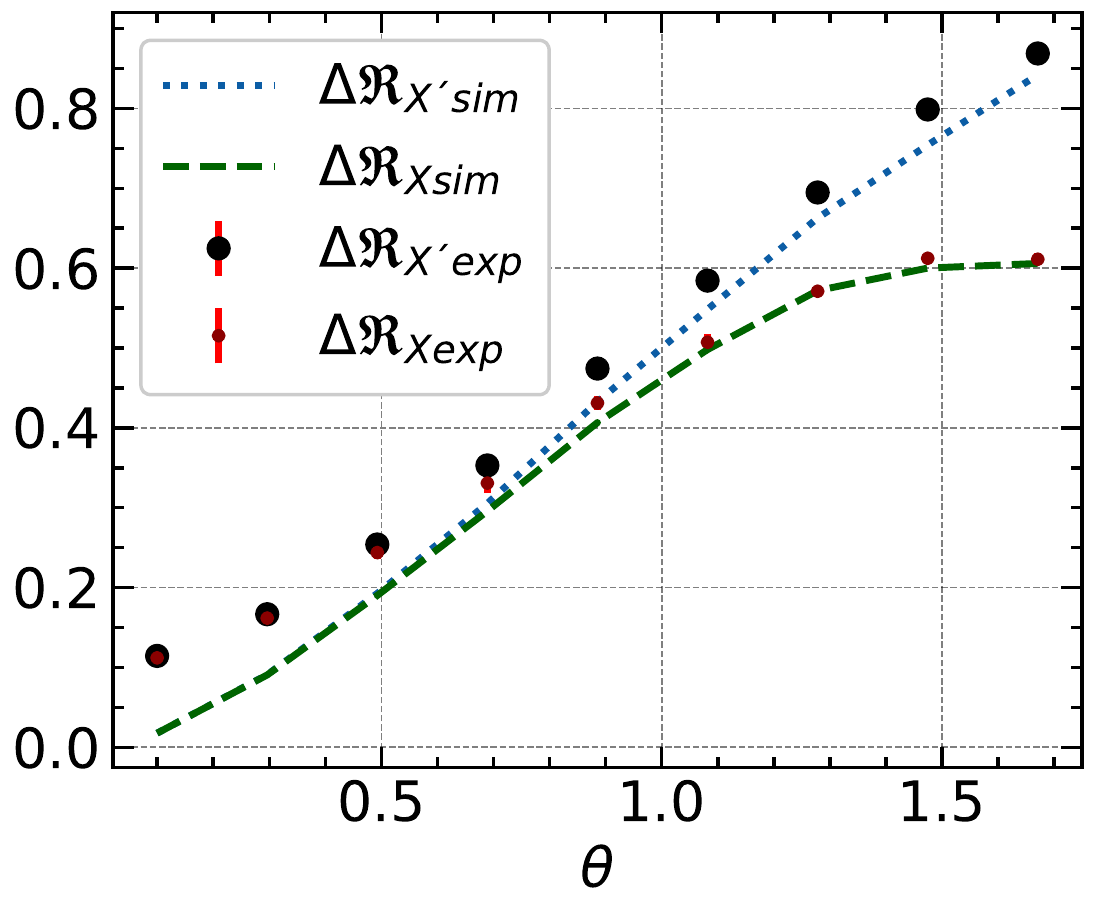}{\label{fig:c}} }}
  \caption{Simulation and experimental results for the comparison between  $\Delta \mathcal{R}_{X'}$ and $\Delta \mathcal{R}_{X}$ in different contexts.}    
\end{figure}

To exemplify the application of this quantum circuit, we explored some of the cases discussed in the last section. We used the Belem quantum chip of IBM's Quantum Experience, whose calibration parameters are shown in Table \ref{tb:belem}. We use two auxiliary qubits, one for implementing $\Phi^{\epsilon}_X$ and the other for implementing $\Phi_{X'}$. In Fig. \ref{fig:a}, we show the experimentally verify results for the verification of the case \textit{(iii)}, where the already established reality of the observable $X'$ is not affected by the monitoring of $X$. To do this, we prepared a qubit in the state $\ket{\psi} = \ket{+} = \frac{1}{\sqrt{2}}(\ket{0} + \ket{1})$ with the monitoring map with regard to the observable $X = \sigma_z$, while the observable $X'$ is $\sigma_x$. In general, the experimental results agree quite well with the theoretical predictions. However, due to hardware errors and decoherence, for small values of $\theta$ we have a small deviation between theory and experiment, and we notice that for some values of $\theta$ we can have 
%$\Delta \mathfrak{R}_{X exp} > \Delta \mathfrak{R}_{X sim}$, while 
$\Delta \mathfrak{R}_{X' exp} < 0$. 
%This is due to the fact that $S(\Phi^{\epsilon}_X(\rho_{exp})) > S(\Phi^{\epsilon}_X(\rho_{sim}))$ which explains $\Delta \mathfrak{R}_{X exp} > \Delta \mathfrak{R}_{X sim}$, $S(\Phi_{X'}\Phi^{\epsilon}_X(\rho_{exp})) > S(\Phi_{X'}\Phi^{\epsilon}_X(\rho_{sim}))$ and $S(\Phi_{X'}\Phi^{\epsilon}_X(\rho_{exp})) > S(\Phi^{\epsilon}_X(\rho_{exp}))$, whereas in this case we should have $S(\Phi^{\epsilon}_X(\rho)) = S(\Phi_{X'} \Phi^{\epsilon}_X(\rho))$ for $\Delta \mathfrak{R}_{X´ exp}$ to be zero. 
In Fig. \ref{fig:b}, we experimental results for the case \textit{(v)}. For this, we prepare the initial state in $\ket{\psi} = \frac{1}{\sqrt{2}}(\ket{0} + i \ket{1})$ with the monitoring map implemented using the observable $X = \sigma_z $, while the observable $X'$ is chosen to be $\sigma_x$. Finally, in Fig. \ref{fig:c}, we experimentally verify the case where $\Delta \mathfrak{R}_{X'} \ge \Delta \mathfrak{R}_{X}$. To do this, we follow the example given in the previous section. We prepare the initial state $\ket{\psi} = \ket{+}$ with the monitoring map giving by $X = \hat{n} \cdot \vec{\sigma}$, for $\theta = \pi/4$, while $X' = \sigma_z$. For these last two examples, we observe that, for small $\theta$, we have $\Delta \mathfrak{R}_{X exp} > \Delta \mathfrak{R}_{X sim}$ and $\Delta \mathfrak{R}_{X' exp} > \Delta \mathfrak{R}_{X' sim}$.

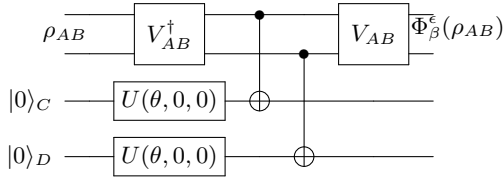
\begin{figure}[t]
%\centering
$
\Qcircuit @C=1em @R=.7em {
& \qw & \multigate{1}{V_{AB}^{\dagger}} & \ctrl{2} & \qw & \multigate{1}{V_{AB}} & \qw \\
\ustick{\rho_{AB}} & \qw & \ghost{{V_{AB}^{\dagger}}} & \qw & \ctrl{2} & \ghost{{V_{AB}}} & \qw  &  \ustick{\Phi^{\epsilon}_{\beta}(\rho_{AB})} \\
\lstick{|0\rangle_{C}}  & \qw & \gate{U(\theta, 0, 0)} & \targ & \qw & \qw  & \qw \\ 
\lstick{|0\rangle_{D}} & \qw & \gate{U(\theta, 0, 0)} & \qw & \targ & \qw  & \qw 
}
$
\caption{Quantum circuit to perform weak non-revealing von Neumann measurements $\Phi^{\epsilon}_{\beta}(\rho_{AB})$ in the general two-qubit basis $\beta = \left\{|n_{j,k}\rangle_{AB}=V_{AB}|j\rangle_{A}\otimes|k\rangle_{B}\right\}_{j,k=0}^{1}$. }
\label{fig:2qbqc}
\end{figure}

Now, for completeness, let us give the quantum circuit for the two-qubit case. In view of the development made previously for the one qubit case, one can see that if we use two auxiliary qubits $C$ and $D$ and two controlled phase-operations $C_Z(A\rightarrow C)$ and $C_Z(B\rightarrow D)$, we can utilize $|\tau\rangle =C_Z(B\rightarrow D)C_Z(A\rightarrow C)|\Upsilon\rangle_{AB}\otimes|\phi, \phi\rangle_{CD}$ to implement the monitoring map in the computational basis: $\Phi^{\epsilon}_{0,1}(|\Upsilon\rangle_{AB}) = Tr_{CD}(|\tau\rangle\langle\tau|)$. Above and hereafter, we use the notation $|jk\rangle\equiv|j\rangle\otimes|k\rangle$. A monitoring map in a general two-qubit basis $\beta = \{|n_{j,k}\rangle = V_{AB}|jk\rangle\}_{j,k=0}^{1}$  can be written as follows $\Phi^{\epsilon}_{\beta}(\rho)= (1 -  \epsilon) \rho + \rho \sum_{j,k}|\beta_{j,k}\rangle\langle\beta_{jk}|\rho|\beta_{j,k}\rangle\langle\beta_{jk}| = V_{AB}\Phi^{\epsilon}_{0,1}(\tilde{\rho})V_{AB}^{\dagger}$, where $\tilde{\rho}=V_{AB}^{\dagger}\rho V_{AB}$. So, the quantum circuit to implement this general two-qubit weak non-selective von Neumann measurement is shown in Fig. \ref{fig:2qbqc}. Finally,  we must mention that the extension of the quantum circuit depicted in Fig. \ref{fig:2qbqc} for $n$ qubits is straightforward. Using $n$ auxiliary qubits and $n$ controlled-phase operations, we can implement the monitoring map in the computational basis. The monitoring map in an arbitrary $n$-qubit basis can then be obtained using the associated $n$-qubit unitary transformation.

%------------------------
%\section{Quantum Erasure}

%---------------------
\section{Conclusions}
\label{sec:con}
In this work, we addressed several interesting questions regarding the variation of the reality of observables under the monitoring map. We learned that the variation of reality is highly dependable on the context, i.e., it depends on the prepared state and on the choice of the observable to be used for implementing the monitoring map. For instance, we showed that if the observables are compatible, then the variations of reality of the two observables are equal. On the other hand, if the observables are maximally incompatible, then the variation of reality of the observable $X$, that is used to apply the monitoring map, is bigger than variation of reality of the other observable $X'$. However, interestingly, the monitoring map of the observable $X$ does not affect the reality of the observable $X'$, when its reality was already established. Besides, the variation of reality of two maximally incompatible observables $X$ and $X'$ can be equal when the prepared state is an eigenstate (or a mixture of eigenstates) of a third observable $X''$ that, by its turn, is also maximally incompatible with the other two observables $X$ and $X'$. Finally, we showed the unexpected case where it is possible that the variation of reality of $X'$ can be bigger than the variation of reality of the observable $X$, under monitoring of $X$. It's also worth mentioning that the results obtained here can be extended for the bipartite (ir)reality defined in \cite{Renato}. Besides, we provided a quantum circuit that can be used to implement such monitoring maps, the weak non-revealed von Neumann measurements, for an arbitrary number of qubits in an arbitrary basis. We then used this quantum circuit to test the variation of reality of observables using the IBM's quantum computers. Our experimental results agreed quite well with our theoretical predictions. Besides the interesting aspects of quantum reality we reported in this article, we expect that the introduced quantum circuits will be very useful for experimental investigations involving monitoring maps.

\begin{acknowledgments}
This work was supported  by Universidade Federal do ABC (UFABC), process 23006.000123/2018-23, and by the Instituto Nacional de Ci\^encia e Tecnologia de Informa\c{c}\~ao Qu\^antica (INCT-IQ), process 465469/2014-0.
\end{acknowledgments}

\vspace{0.3cm}

\textbf{Data availability} The Qiskit code used for implementing the experiments to obtain the data used in this article is available upon request to the authors.

%--------------------

\end{document}